# Finding balance between work and play

In 2020 we all had to rethink our working spaces and schedules, testing the boundaries between work and home—and we saw that line can be hard to draw.

01

# Finding balance between work and play

**In this report, we investigate people's work patterns over the previous year and how they impacted productivity and developer experience.**

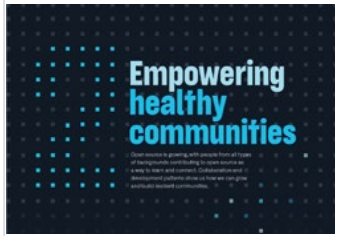

Empowering communities

**Productivity report** ➔

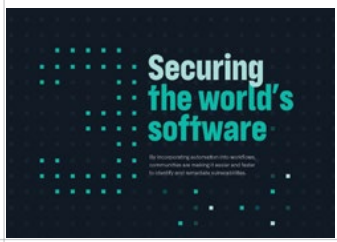

Securing software

**Security report** ➔

//table of contents

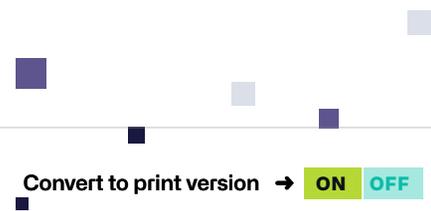

Convert to print version ➔ ON OFF

# Executive summary

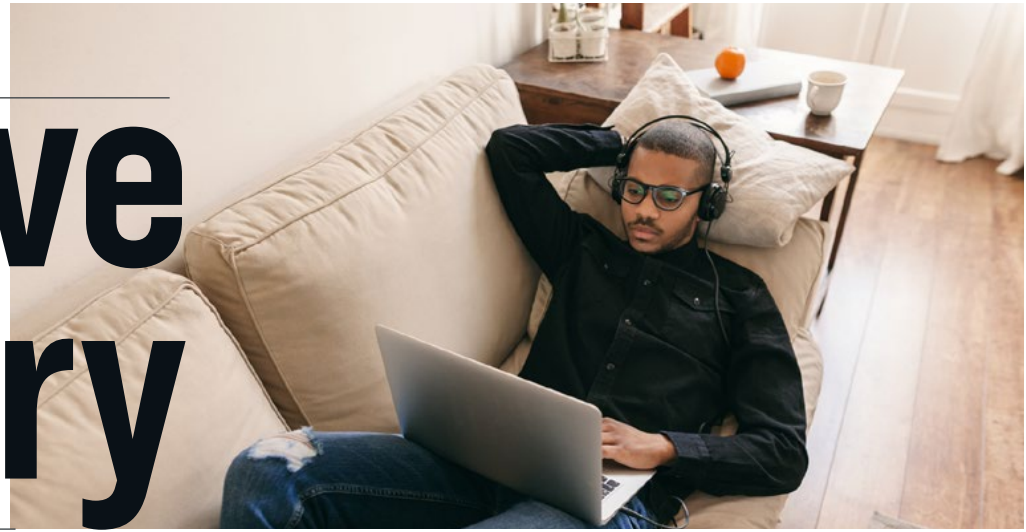

Over the past year, many developers and other technology professionals have transitioned to a remote-first world, as COVID-19 pressed organizations to support working from home whenever possible. This shift quickly changed the routines and environments where we work and learn, redrawing the lines between personal and professional lives. How does this affect the ways we develop and deliver software, both at work and in our open source projects?

## 34%
more pull requests when teams automate workflows

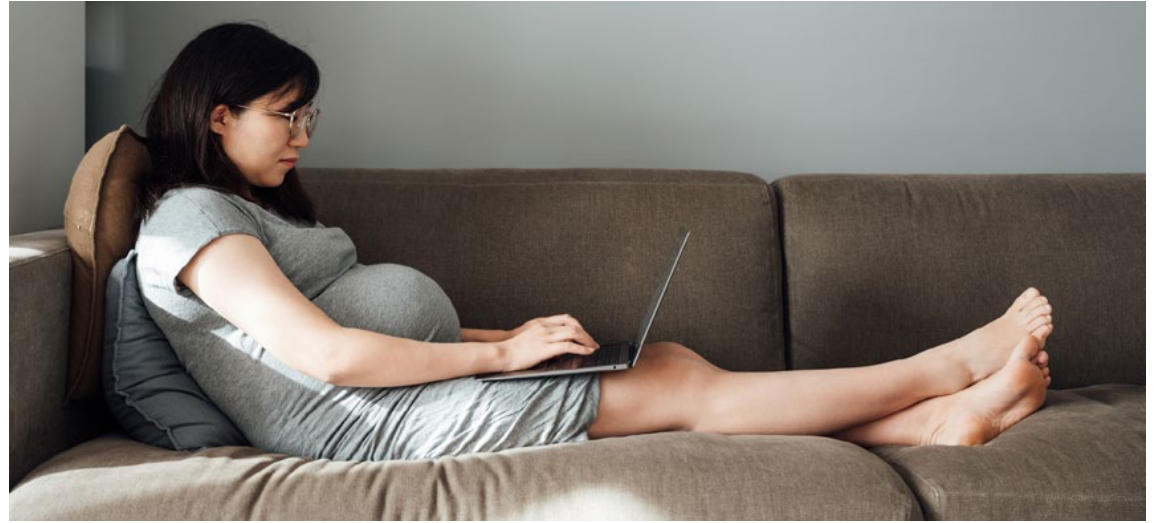

# 18 %

faster pull request merges when teams automate

Here we investigate two aspects of developer productivity that may be affected by this shift to work from home: work rhythms and activity. We build on our previous Octoverse Spotlight analysis by expanding the time horizon, investigating different time zones, and adding benchmarks for developer activity. In our findings, we share insights for software developers and company leaders who are guiding newly distributed teams through uncertainty.

As the largest global developer platform, GitHub is in a unique position to see patterns and changes in developer activity. Understanding a new work environment for developer workflows can help us all adapt to our changing remote-first world, and prepare for the future of work.

# Key findings

**01**

### Small pull requests drive innovation and productivity.

Teams that focus on small pull requests and closer collaboration have better reviews and faster feedback. Throughout the year, developers stepped up their work by keeping pull requests at the same size or smaller and merged pull requests up to seven and a half hours faster. This gives developers more time to do the things they love.

**02**

### Automation drives productivity gains and improves developer experience.

Open source repositories that use Actions to automate pull requests—a key stage in software workflows—see time to merge decrease by 18% and the number of pull requests merged increase by 34%. By leveraging automation in their workflows, teams minimize manual work and reclaim time for innovation, development, and collaboration.

**03**

### Open source is a great escape when everyone is stuck at home

Analysis shows that developers "walk away" from their work on holidays and weekends, while open source projects see activity spike during those times. This suggests that open source is viewed differently from other work and may be an outlet, providing a great opportunity for people to learn, grow, be creative, and engage with community.

**04**

### Developer activity highlights the importance of flexibility and personal solutions.

We see increased development work—both time spent and amount of work—across all time zones we investigated. Developers may be taking advantage of flexible schedules to manage their time and energy, which contributes to this sustained productivity. We caution that if work happens at the expense of personal time and breaks to maintain a healthy work life balance, this pace may not be sustainable in the long run.

# Take action

Take these actions to find balance and create a better developer experience.

**01**

### Manage your energy.
One of the best ways to manage working from home is to think about managing your energy: Plan focus time for core work, take breaks, and use low-energy time for meetings. This can help counteract long days that can make work feel repetitive and draining.

**02**

### Be flexible to find unique solutions for the best outcomes.
People can be resilient. When you allow flexible schedules, your team members can optimize their days, manage their energy, and meet work demands in challenging times. Organizations that can adapt quickly will also drive innovation as you face new challenges and opportunities. Flexible tools and processes let your teams plan, track, and develop software regardless of where they work, and shifting tools to the cloud provides a superior developer experience.

**03**

### Use automation and small pull requests to ship better code faster.
Teams can develop more effectively and efficiently by creating smaller pull requests and automating the development workflow. This allows you to collaborate more efficiently, conduct better code reviews, and reduce manual work so you can ship code faster. Taking these steps gives your team more time for future development and peer review, so you can ship even more features, faster.

**04**

### Embrace and support collaboration and open source.
People are turning to open source as an outlet for creativity and learning—finding ways to mentally "step away" from your work and then "step into" creative or learning projects—even if they happen on the same computer screen. Examine your policies around external work and technology to ensure they allow time for learning, including moonlighting policies that allow for external projects. Recognizing that open source is a platform for more than just work is important for employees' wellbeing.

# When and how much we work

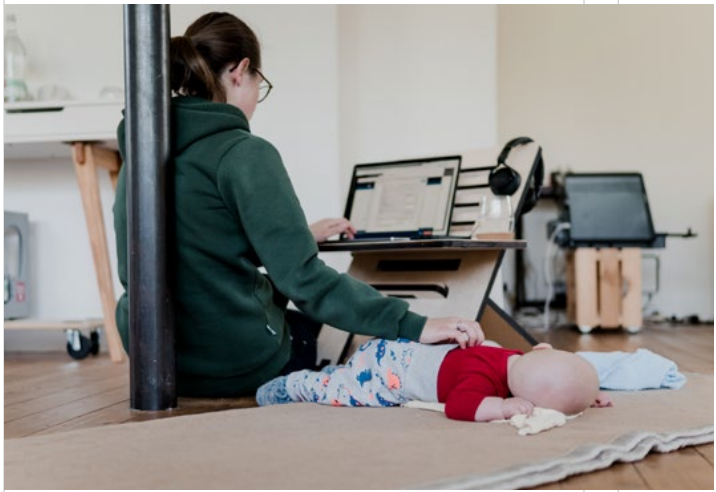

D eveloper productivity has several dimensions, including the ability to do complex work, find "flow," and complete work. One aspect of developer productivity is the span or timing of work. Developers report having a "most productive time of day," and some stretch out their work over time, depending on what works best for them (Meyer et al. 2017). While work span isn't the only component of productivity, when paired with the work that is done, it can give us insights into how our days are structured.

## Data for this section

The data for this section of the report comes from analyzing paid organization accounts that meet the following criteria:

- Created before October 1, 2018 with activity each month through September 2020
- On a paid Team or Enterprise Cloud account

To allow for easier year-over-year comparisons, we normalize our analysis using per-user figures, unless noted otherwise. Only aggregate, anonymous data is reported.

More than 35,000[1] organizations are included in our analysis, with strongest representation in North America (41%), Europe (35%), and Asia (17%). Thus, our analysis investigates developer work patterns for four time zones around the world: UK Time Zone, US Eastern Time Zone, US Pacific Time Zone, and Japan Standard Time Zone.

### Distribution of active users by time zone

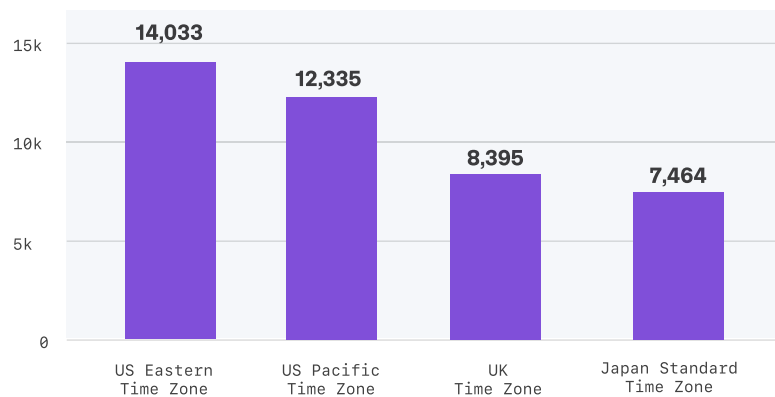

| Time Zone | Users |
| --- | --- |
| US Eastern Time Zone | 14,033 |
| US Pacific Time Zone | 12,335 |
| UK Time Zone | 8,395 |
| Japan Standard Time Zone | 7,464 |

### Development doesn't change

Over the last year, COVID-19 changed how and where people work but development itself doesn't change. The best way to investigate developer productivity during changes in context and environment is to use a measure that doesn't change when work does. To find a robust measure, we can ask people directly (in interviews or surveys) or look at artifacts created when they do their work (like notebooks, source code, or system logs).

Development activity on GitHub is a good proxy for activity that is fairly robust to shifting work routines. While some things like kanban boards may have moved from whiteboards to online tools, development can be observed via the same pushes, pull requests, and issues regardless of whether we do that work from an office or at home. This means our year-over-year comparisons are reliable.

[1] This is less than were included in our Octoverse Spotlight because this analysis required contributions across 24 months instead of 15 months previously.

GitHub data can't capture if parts of the day feel more or less productive to developers, but we can observe patterns in work activity and how these patterns change over time. These patterns in development work aren't a complete picture of a developer's day—they simply show us time spent on development work. We know days are also filled with meetings, planning, and email. However, these patterns may give us some insights into work and how it may have changed, particularly during a large shift to work from home.

People who typically work from home tend to work more hours—up to one or two eight-hour days more per week ([Hill et al. 2010](#)). This is likely because our work stretches into our lives, and boundaries between work and home blur. With the sudden shift to remote work, we wondered if we would see any patterns of increased development time.

Within those time zones, we looked at both development windows and work volume. We selected these time zones because they had strong shelter-in-place responses, which may help highlight any resulting changes to work patterns. These were also chosen because our sample size was large enough to not be overly influenced by a few organizations and allow for significant results.

The work day for an individual developer was captured as the time between their first and last `git push` to any repository's default branch—a "push window." This is a rough paradigm for when someone's development window begins and ends.

During this time frame, we measured work volume by number of pushes. Note this only captures development work, and development is only part of a developer's day; many other work tasks such as planning, designing, meetings, and email are important and can happen outside of this window.

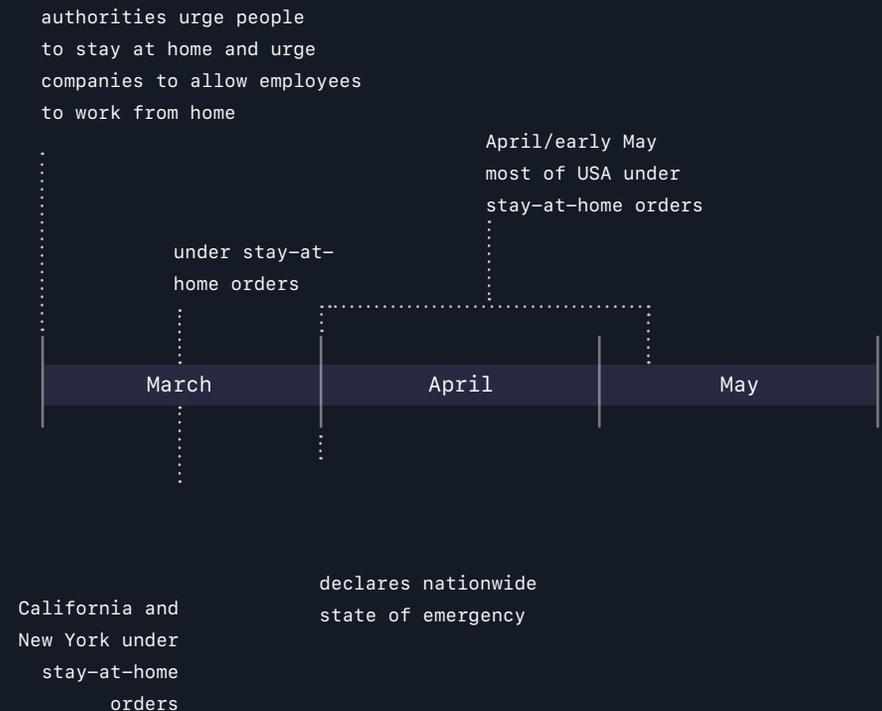

**Key Dates**
Work has changed for many developers due to the COVID-19 pandemic and shifts to working from home. We summarize notable dates here.

- authorities urge people to stay at home and urge companies to allow employees to work from home
- under stay-at-home orders
- April/early May most of USA under stay-at-home orders

March — April — May

- California and New York under stay-at-home orders
- declares nationwide state of emergency

## Global work at-a-glance

Comparing work span across all time zones we found:

- Mondays have a shorter push window in all time zones, with most working 253 minutes (UK) to 284 minutes (Japan), or 4.2 to 4.7 hours respectively.

- All time zones work on Saturdays and Sundays, and almost as much as on Mondays. On Saturdays, this ranges from 217 minutes (UK) to 256 minutes (US Eastern), or 3.7 hours and 4.3 hours respectively. And on Sundays, this ranges from 221 minutes (UK) to 320 minutes (US Pacific), or 3.7 hours and 5.4 hours respectively. This may be due to people feeling like they have to catch up on work from the previous week or prepare for the upcoming week, or it could be users spending time on personal projects.

- US Pacific Time Zone sees a marked increase in push window compared to other time zones on Tuesdays through Fridays, ranging from 479 minutes (Fridays) to 494 minutes (Wednesdays), or eight hours and 8.3 hours respectively.

This is a rough approximation of workdays and only captures the span of time spent on development work. Many developers likely spread out their coding and spend time on other tasks: previous research shows that a typical developer's day is spent in several tasks including coding, emails, work-related web browsing, reviewing code, and collaboration. Another study of almost 6,000 developers finds that developers' days have similar tasks yet differ depending on the phase of the software project they are working on. Depending on how developers structure their day, much of their work will happen outside of the push window we have captured.

**Average push window by day of the week, all time zones**

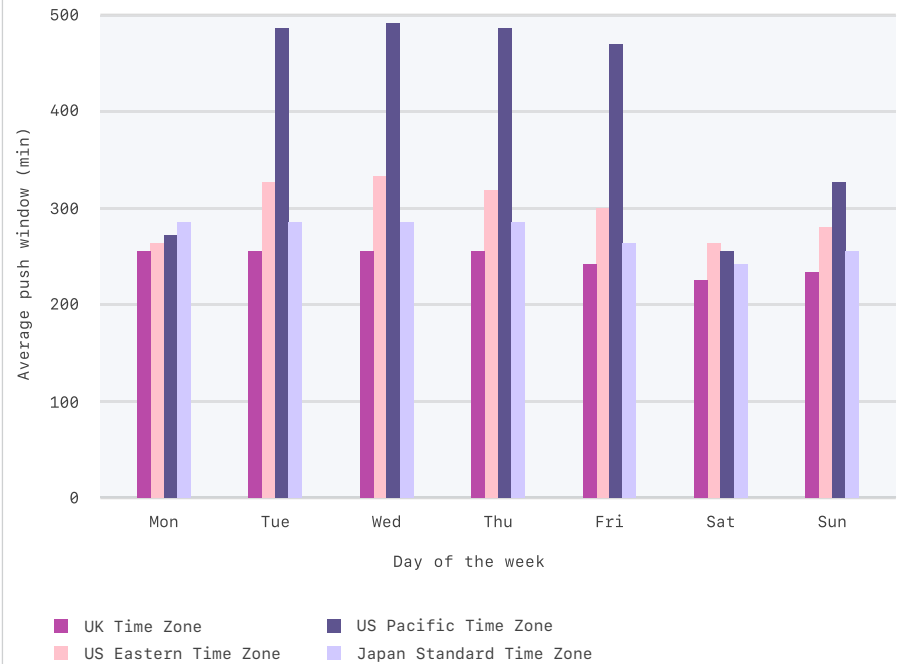

Aggregate findings for the push volume for each time zone also highlight interesting patterns when we display them side by side:

- We see activity seven days a week from developers in all time zones we studied.

- Of the time zones, we see the most balanced activity per person in Japan Standard time zone, which is the most sustainable approach to work. We also note this time zone had a more moderate governmental response to COVID-19, where shifts to home happened early but were not mandatory, so shifts in work routines may not have been as disruptive. Japan also had one of the earliest and best responses to COVID-19, with relatively quick return-to-work scenarios.

- In contrast to the Japan Standard Time Zone, we observe the highest activity per person in the US Pacific Time Zone. This may be driven by tech's culture of overwork (with two strong tech centers located in this time zone), or trying to work across several time zones with other colleagues, stretching from Europe to Asia. This could be concerning for long-term sustainability and burnout.

Next we will examine these time zones in detail, starting with UK Time Zone.

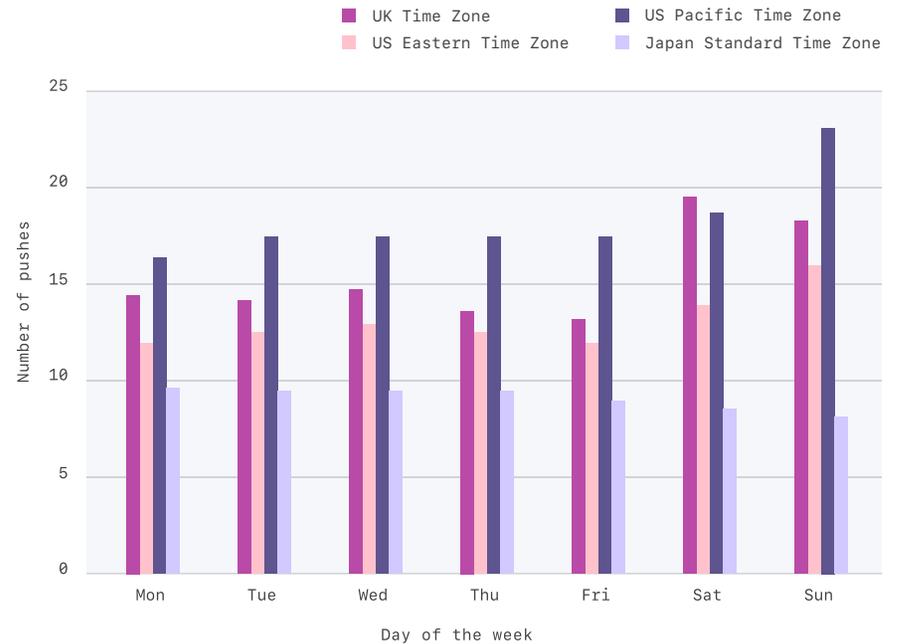

**Average push volume by day of the week, all time zones**

# Push window and work volume: UK Time Zone

We begin our analysis in the United Kingdom, where the first shelter-in-place orders occurred. Here, developers had shorter push windows prior to April 2020 (relative to the same week in the year prior), and then a strong increase starting in mid-March. Push windows start to level off to three minutes longer than the previous year through August.

When looking at work volume, users in the UK Time Zone have a push volume that does not begin to increase until mid-June. It then remains elevated compared to the previous year through August and September. That is, people in the UK Time Zone are working longer hours and also doing more starting in June.

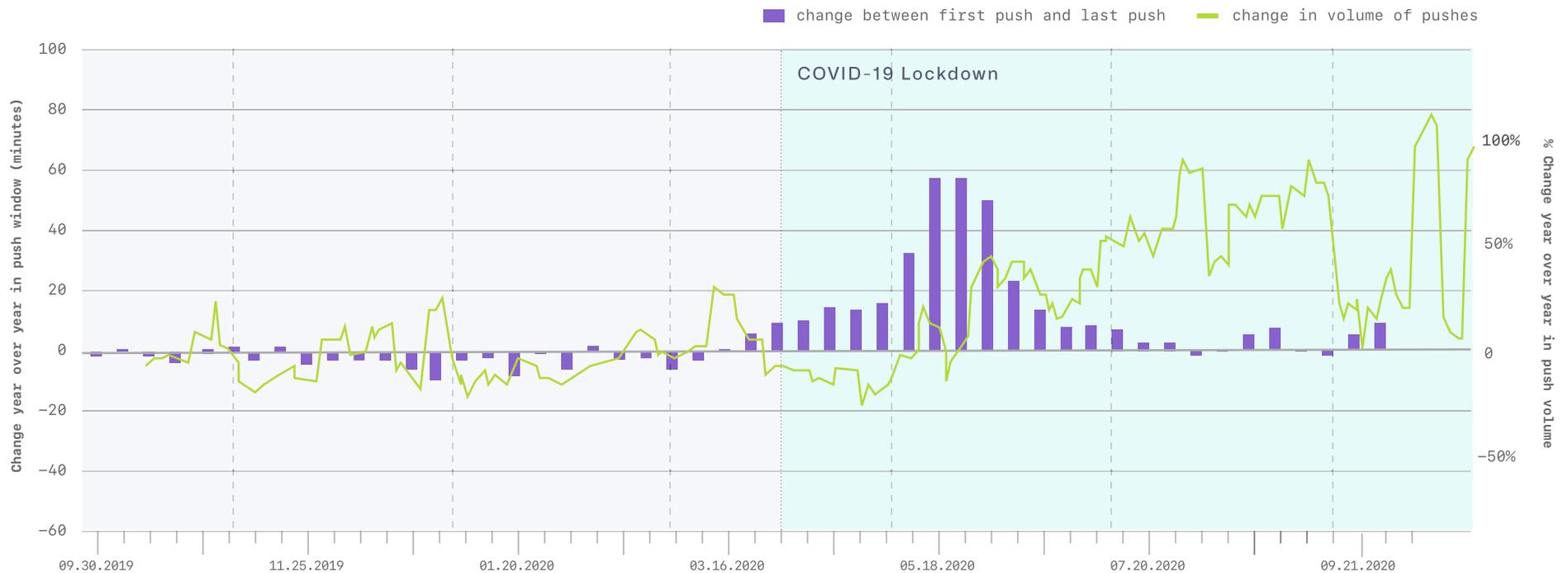

**Change year over year: push window and work volume**
UK Time Zone

An aggregate summary of work across the week in the UK Time Zone shows the average push window span and work volume (in average number of pushes). In this time zone:

- Typical push window ranges from 236 minutes on Friday (3.9 hours) to 261 minutes on Wednesday (4.4 hours).

- Work volume ranges from 13 commits on Friday to 15 commits on Wednesday.

- Commits increase to a high of 20 on Saturdays.

The increased push volume on weekends is likely due to a drop in the number of developers on the weekends. It may also represent an increase in the amount of personal work, such as open source, hobbies, and education. This makes sense, as developers whose weekdays are filled with meetings and other priorities use weekends to make progress on their personal projects. We note this pattern continues through all time zones except Japan Standard. We address this later when we investigate that time zone.

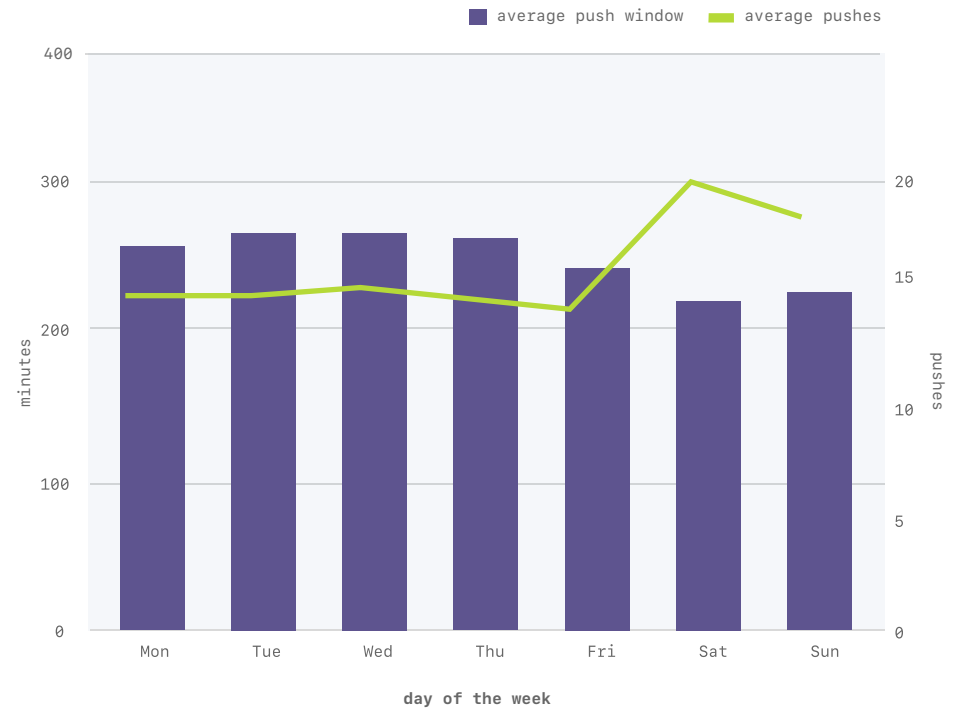

**Average work span and work volume by day of the week**
UK Time Zone

# Push window and work volume: US Eastern Time Zone

Next, we investigate the US East Coast where developers saw shorter push windows for much of 2020, followed by an increase in mid-March, when they leveled off to a two- to three-minute increase in August. The spike and fall in November is due to the shift in the Thanksgiving holiday.

Developers in the East Coast Time Zone have a push volume that is largely steady around 25% to 30% increase compared to the previous year until mid-April, when work volume levels out for a few weeks. We observe a bump in work volume mid-May through late June, which remains even through July and August, then dips in September. This suggests that users are distributing work across longer time periods, particularly as they work from home for an extended period and find flexible solutions to meet their new work needs.

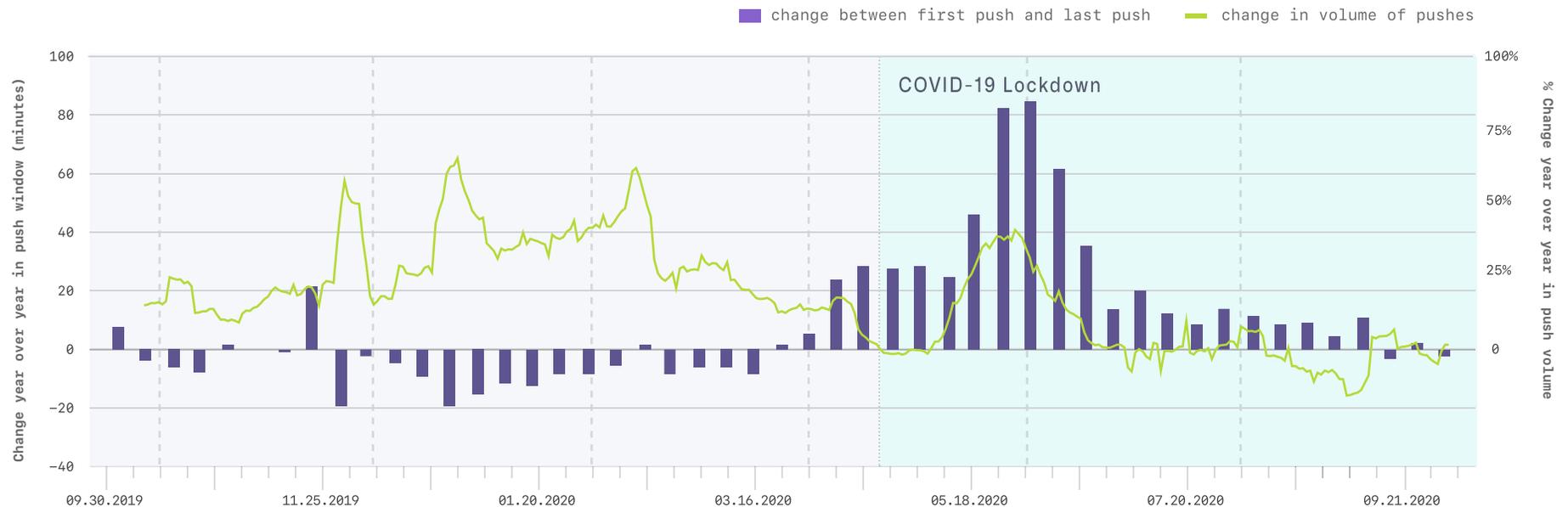

**Change year over year: push window and work volume**
US Eastern Time Zone

The aggregate summary of work across the week in the US Eastern Time Zone represents a typical week spent on development there:

- A weekday push window ranges from 261 minutes on Monday (4.4 hours) to 317 minutes on Wednesday (5.3 hours).

- Work volume ranges from 12 commits on Mondays to 13 commits on Wednesdays.

- Commits increase to a high of 16 on Sundays.

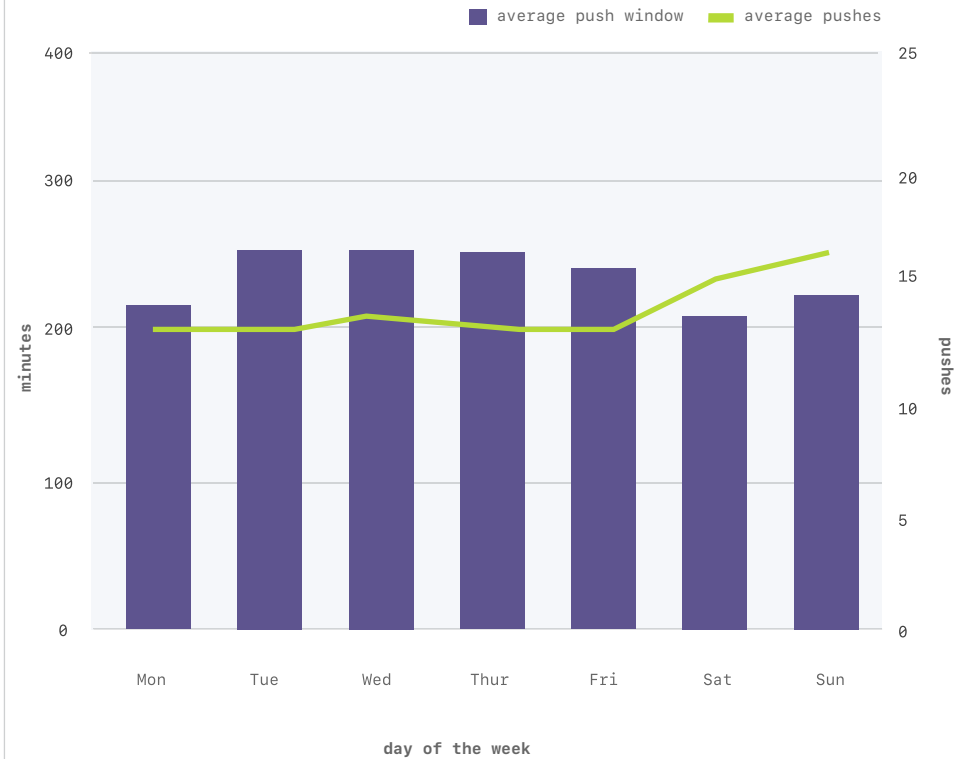

**Average work span and push volume by day of the week**
US Eastern Time Zone

**People are working as much or more on Sundays as on Mondays.**

# Push window and work volume: US Pacific Time Zone

In this analysis, push windows show variable lengths compared to last year and then increase starting mid-March (typically 30 to 60 minutes per day). This pattern holds through June, leveling off in early July. It then shows moderate year-over-year increases through the rest of July (five to seven minutes) until it drops again in August. Note the spike and fall in November is due to the shift in the Thanksgiving holiday.

Developers in the Pacific Time Zone have a push volume that is consistently higher than the previous year. It increases in May, with activity exceeding 50% higher year over year at many points, before falling to previous levels of about a 25% increase. Compared to those in other time zones, these developers are continuing to do the most work in terms of code pushes through the year we investigated.

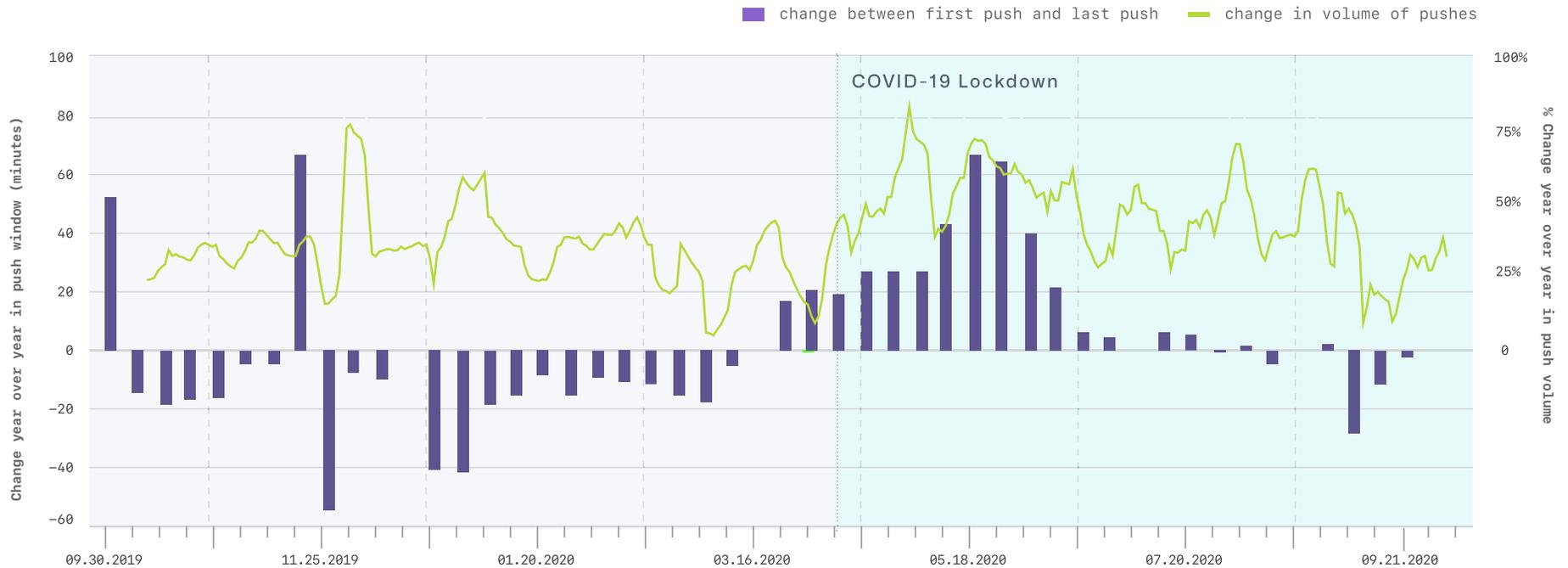

Change year over year: push window and work volume
US Pacific Time Zone

In this time zone:

- A weekday push window ranges from 269 minutes on Monday (4.5 hours) to 494 minutes on Wednesday (8.3 hours).

- Work volume during the week ranges from 16 commits on Mondays to 17 commits on Wednesdays.

- Push windows increase to a high of 23 commits on Sundays.

## Down time is for open source

We see an interesting trend: Enterprise developer activity drops on weekends and holidays. At the same time, open source activity jumps on weekends and holidays—evidence that as people are "signing off" of work, they are "signing on" to open source. Open source project creation is also up by 25% since April year over year.

This is exciting, especially as so many of us are at home right now. Open source gives us an opportunity to make and create, to learn and collaborate, and to share with our community.

Organizations should examine policies around external work and technology to ensure they allow time to engage with learning programs and that moonlighting policies allow for external projects. Recognizing that open source is a platform for more than just work is important for employees' wellbeing.

**Average work span and push volume by day of the week**
US Pacific Time Zone

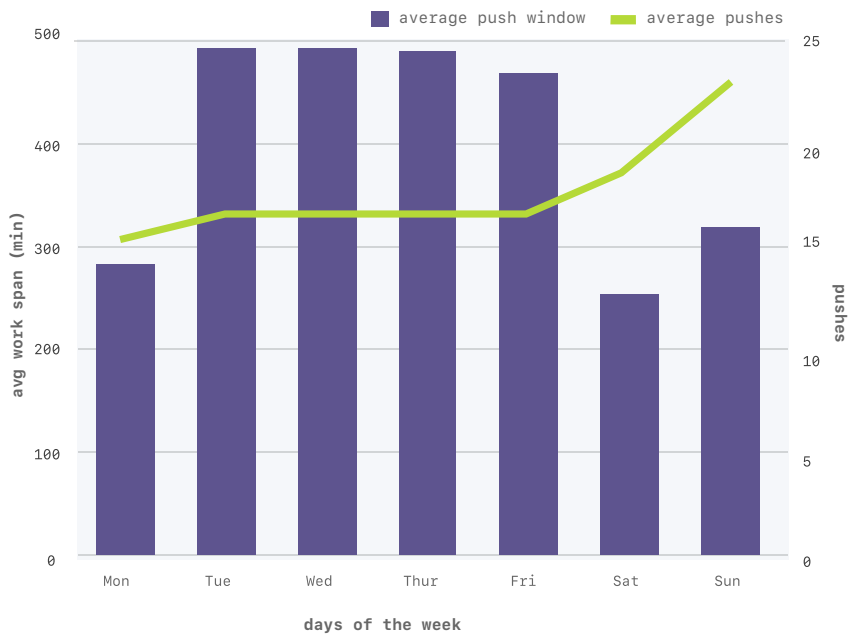

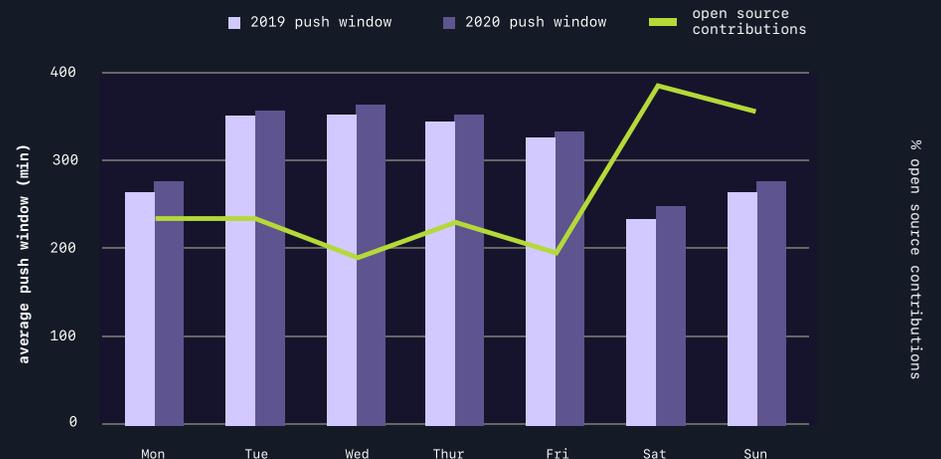

# Push window and work volume: Japan Standard Time Zone

We conclude with the Japan Standard Time Zone. Beginning mid-November 2019, developers in Japan begin working longer days compared to the previous year (at five to 10 minutes longer). Starting in April, we see push windows get longer, with workers spending an additional 20 to 52 minutes per day on work. In June, this dropped back to about 15 more minutes per day than the previous year.

Looking at work volume, developers in the Japan Standard Time Zone have a push volume that remains elevated through April, falling to even in May. In June, individual push volume spikes to 40% to 50% above that of the previous year, before falling to even in July.

**Change year over year: push window and work volume,**
Japan Standard Time Zone

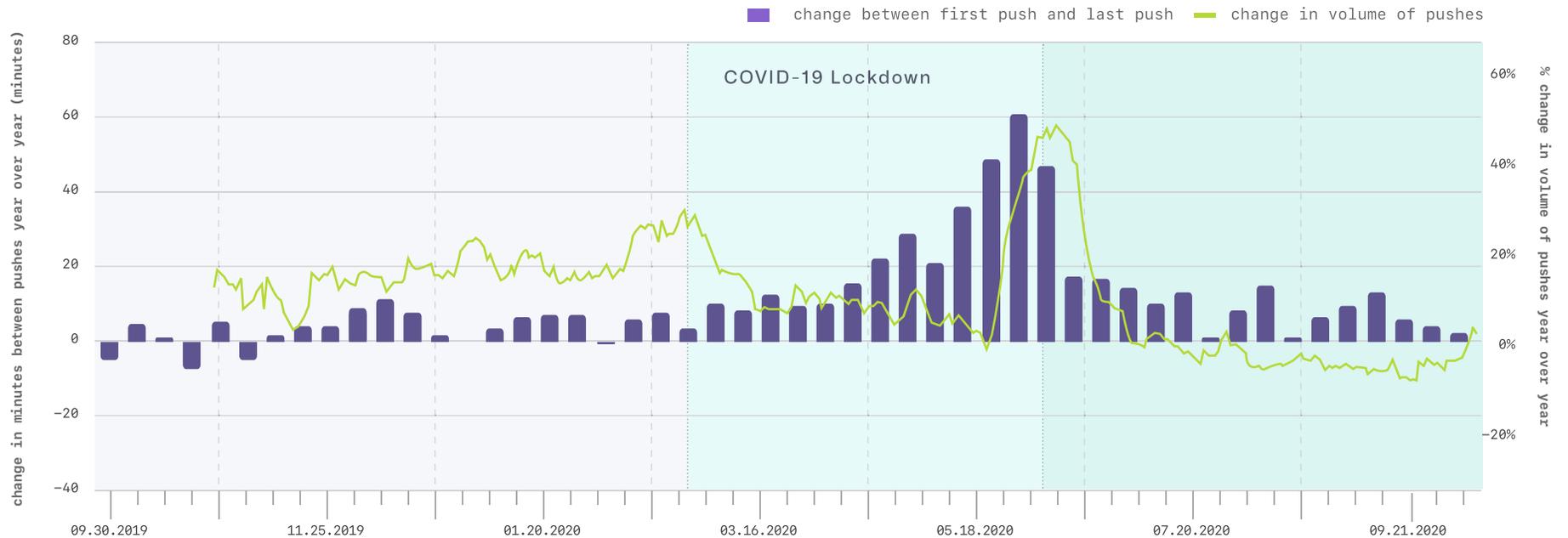

The aggregate summary of work across the week shows us:

- A typical work week push window ranges from 256 minutes on Friday (4.3 hours) to 290 minutes on Tuesday (4.8 hours).

- Work volume during the week has a median of nine commits per developer each day.

- Work volume falls to eight commits on Sundays.

Unlike the shelter-in-place and similar lockdown measures that were implemented in the US and the UK, the measures available to prefectural governors in Japan are non-compulsory in nature, and there are no penalties for individuals or businesses. Therefore, any effects from a shift in work patterns are likely muted.

**Flexibility is important as people work longer hours and do more work.**

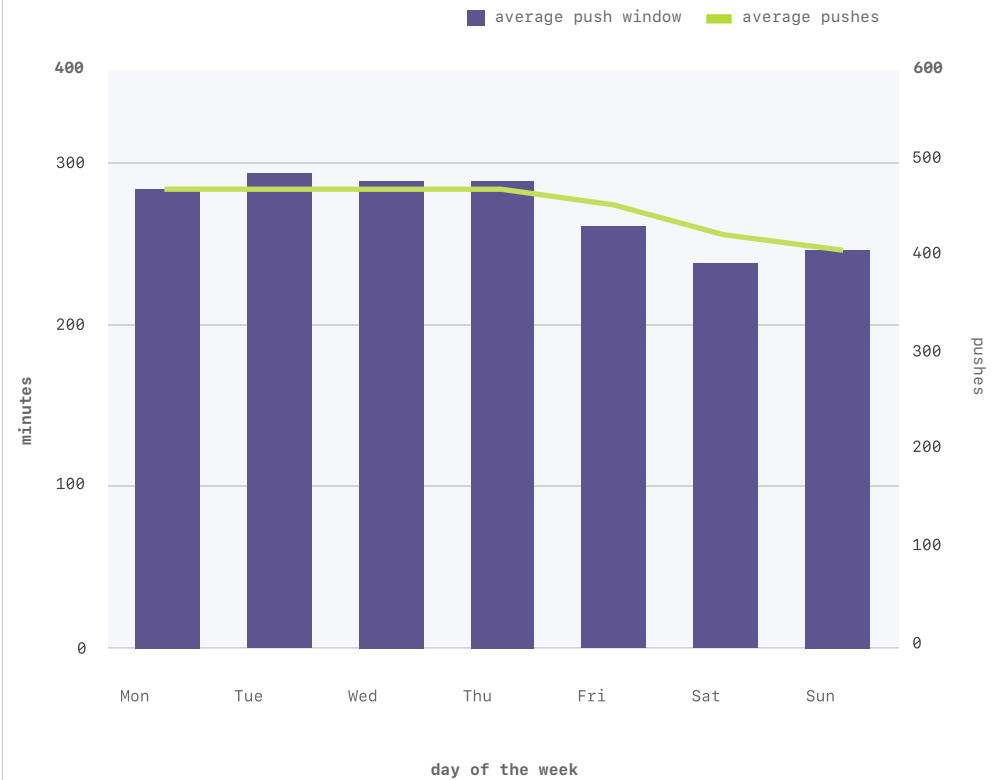

**Average work span and push volume by day of the week**
Japan Standard Time Zone

## Productivity is personal

COVID-19 and the sudden shift to working from home is the ultimate disruption. And yet, developers are working even more than this time last year. We cheer this as evidence of ongoing productivity in the face of uncertainty, but just how sustainable is this?

For some, the shift to working from home has unlocked personal productivity. They can set up their work environment as they'd like, minimize distractions, and have a flexible schedule—perhaps sneaking in a midday nap, a workout, or spending more time with their family. These accommodations weren't possible before because their open office plans were too noisy, or their in-office environment and commutes didn't provide the flexibility to schedule their day for their very best work.

But for every person thriving in our new work-from-home world, there is another colleague struggling.

For some, the shift to work from home means a loss of essential connection with their peers, difficulty communicating, and disruptions. Many people were not set up to work from home, lacking workstations, workspaces, and adequate internet. Many also need to provide childcare and homeschooling for children who are also now shifted to home. Any time reclaimed from no longer commuting has been absorbed by work that grows to fill the space, as the boundary between work and life is blurred. These developers have less time to complete their work, and they feel it in their days.

Tips for improving productivity and avoiding burnout:

- Take a few minutes each day to reflect on something you're grateful for. Some developers report that this has a positive impact on their frame of mind.

- Instead of managing your time, manage your energy. Identify patterns that help you maintain higher levels of energy, and optimize for those. If you're a morning person, get your important tasks done then. If you hit your stride in the late afternoon or evening, see if you can arrange with your team to take a later shift.

- Support flexible, sustainable work schedules and watch for signs of burnout in team members. This helps keep our teams and ourselves happier and more productive.

### For more tips on working from home, check out these resources:

→ **Remote work: How parents are adapting and  working during COVID-19**

→ **Remote work: Working together when we're not together**

→ **Parent-driven development podcast**

The research for this sidebar comes from our colleagues in the SAINTes group:
- Butler and Jaffe (2020) Challenges and Gratitude: A Diary Study of Software Engineers Working From Home During Covid-19 Pandemic
- Ford et al. (2020) A Tale of Two Cities: Software Developers Working from Home During the COVID-19 Pandemic

# Developer activity

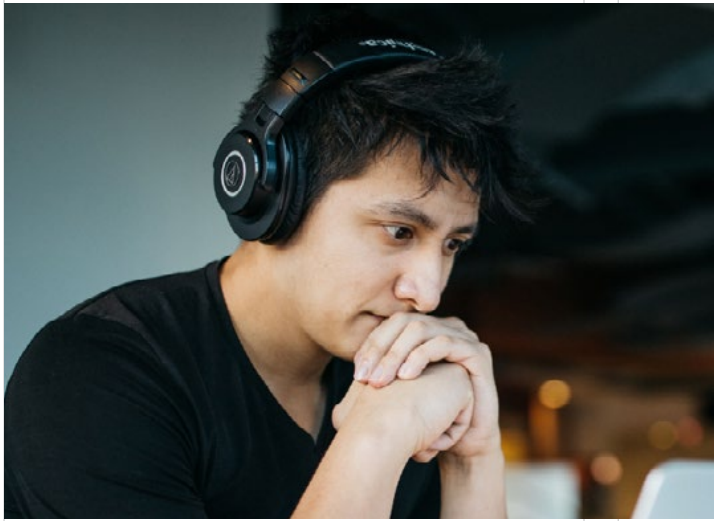

**D**eveloper activity is another aspect of productivity. Measuring developer activity as a component of productivity is complex but rewarding when it's done right. For developers, it can reveal best practices for task management, coordination, and problem solving. And for team leaders, it can remove barriers, help teams work better together, and improve outcomes. Our analysis includes several measures of developer activity, informed by prior research on developer productivity (Ford et al. 2020, Meyer et al. 2014).

## Data for this section

The data for this section comes from analyzing all GitHub activity—public (including open source) and private—year over year. The period of comparison is October 1, 2019 through September 30, 2020 vs. October 1, 2018 through September 30, 2019. The change in the geographic distribution of active users included in the analysis year over year is shown in the chart.

**Geographic distribution of active users**
Current and absolute change relative to 2019

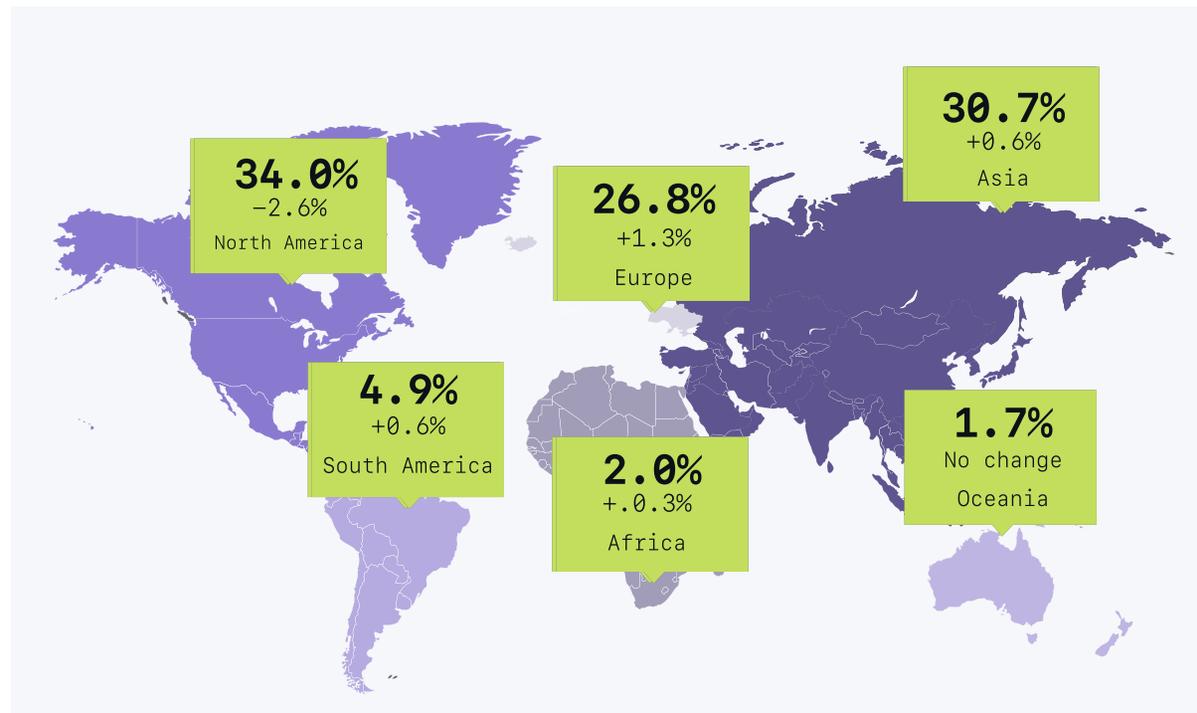

## Increased activity through pandemic

For our analysis of developer productivity, we investigated pull requests, pushes, reviewed pull requests, and commented issues per person. Overall, we see consistent or increased activity for these measures compared to last year.

**Daily active users compared year over year**

Note that as holidays such as Thanksgiving and Lunar New Year occur on slightly different dates in different years, the dips in activity they cause show up offset when we compare years

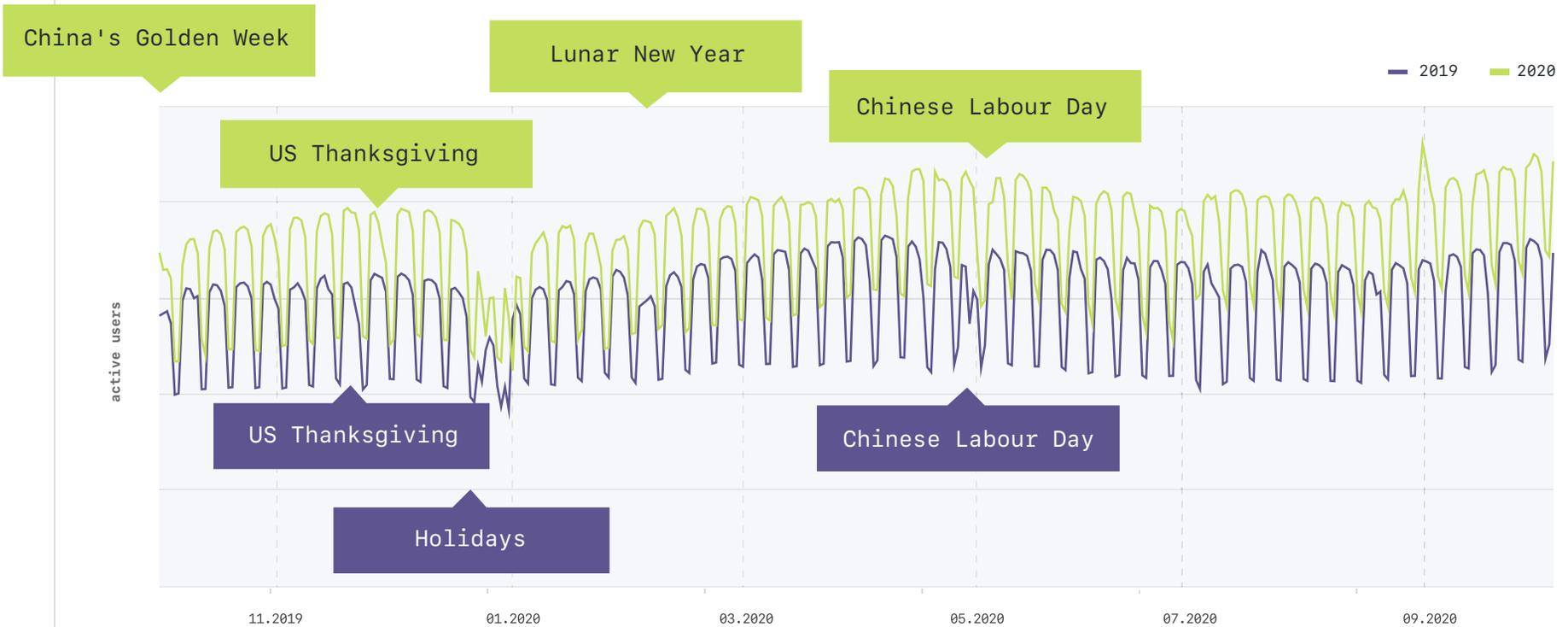

Charts are shown with normalized, per-person activity with a seven-day rolling average for readability, except where otherwise noted. The dip in late 2019 corresponds to the holidays.

We analyzed by pull requests and push volume per person per day, and see that activity is consistently up compared to last year. For those who are curious, detailed charts are included [in the appendix](#).

We also investigated reviewed pull requests and commented issues, and the pattern is similar: higher than last year and consistent through the year. We have not included the reviewed pull requests or commented issues charts for the sake of brevity.

The data provides additional evidence that activity has stayed consistent and even increased throughout the pandemic and the shift to working from home. This is notable because sustained activity through large shifts in how we work show that flexible tools, processes, and solutions can support developer productivity and even continued innovation in the face of disruption. Cloud-based development and a focus on the developer experience supports more stable and resilient development for people, teams, and organizations. Adding to this, our earlier analysis shows that developers benefit from flexibility that allows them to do their work by spreading it out.

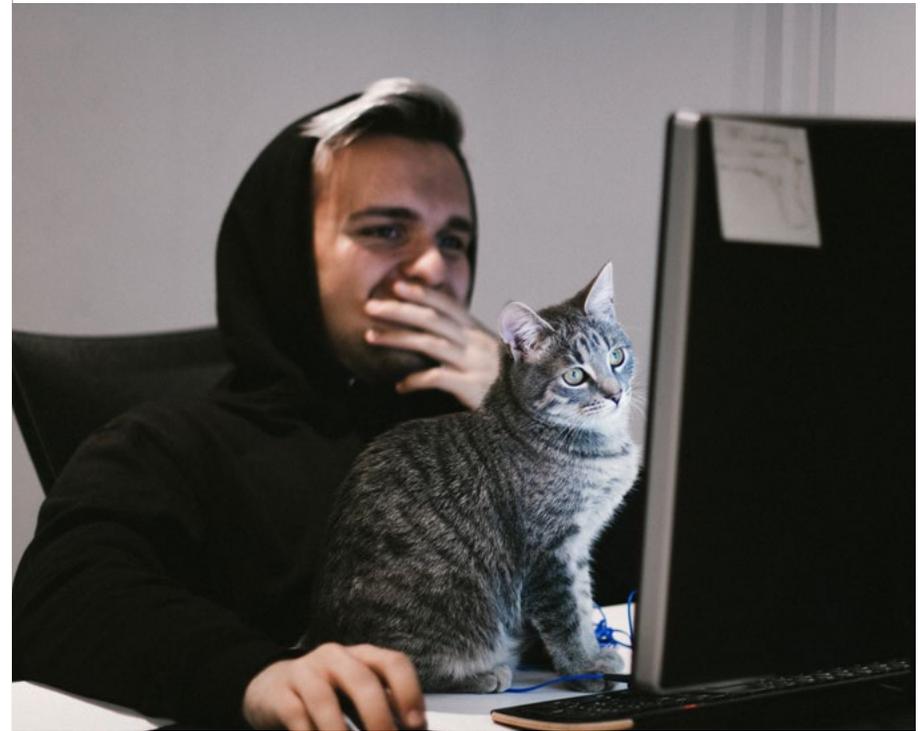

# Development is about collaboration

While some teams have always worked remotely—including both full-time and open source teams—others had to start working from home for the first time. We wondered how the changing environment may have impacted a key piece of collaboration: peer review through pull requests.

A pull request is how developers tell others about changes they make to a repository. Merging a pull request involves a group of interested developers reviewing changes, discussing code, and sometimes following up through commits. Finally, the pull request is merged into the relevant branch of the intended repository. As a proxy for this collaborative process, we measured time to merge—how long teams take to merge pull requests—compared to last year.

In open source repositories, the time to merge a pull request varies through the end of 2019, with the majority of big fluctuations happening around the holidays—shifting from 3.5 hours slower to three hours faster than the previous year for typical swings. However, in mid-February 2020, the time to merge a pull request gets faster than the year prior and stays faster, with times ranging from one to seven hours faster.

**Year-over-year change in time to merge pull requests for open source projects**

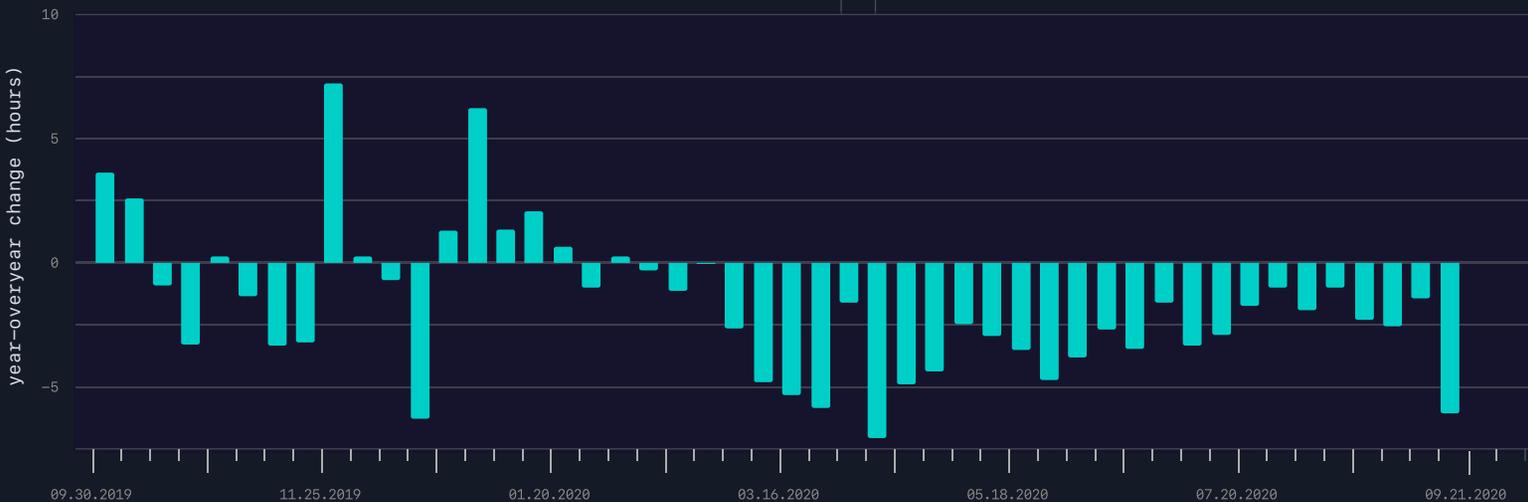

Note: Data for this chart comes from the source used for the Productivity and activity section of this report

In work contexts, the average time to merge pull requests is higher at the end of 2019 than the previous year, at 1.5 hours longer for Team repositories and 1.5 to 4.2 hours longer for Enterprise accounts. Merge times rise over the holidays as team members take time off and bandwidth to review pull requests decreases.

In early 2020, we see longer times than the previous year, though somewhat improved: one to two hours longer for Team repositories and 0.1 to one hour longer for Enterprise Cloud repositories. In March, the average time to merge pull requests drops after a short surge, and holds that pattern through the rest of the year: Team repositories merge pull requests up to five hours faster, and Enterprise Cloud repositories up to six hours faster.

In a recent series of interviews, developers and engineering leaders shared how their teams use pull requests. The number one best practice identified was keeping pull requests small, because it makes reviews easier, leads to better reviews, and makes it easier to revert if there are issues. It also streamlines feedback, creating momentum and contributing to the team's productivity.

"Most of my work as an engineering manager has been teaching people how to scope work to smaller bits. Any time you have a big honking PR it's a nightmare to review either too much discussion or none because no one can get their head around it."

— Sarah Drasner, VP Developer Experience at Netlify

For more on how to break down work and keep pull requests small, check out "How to Scope Down PRs" from Sarah Drasner.

**Year-over-year change in time to merge pull requests for Enterprise Cloud and Team repositories**

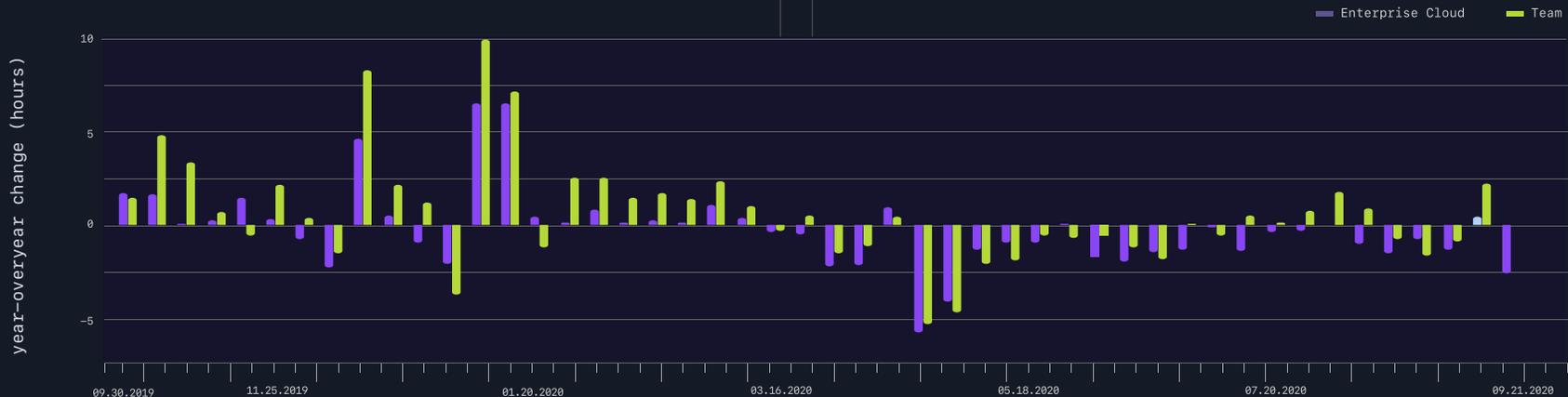

# Distributed work causes flux: issues created per person

We also measure developer activity by looking at the number of GitHub issues created, and see different activity patterns when compared to pull requests and pushes. When compared to the year before, the number of issues created per day on GitHub is less than or equal to what it was before the COVID-19 outbreak. This started to shift in mid-March and continues throughout the period of analysis, as noted by the arrow in the chart. Again, the dip in activity at the end of 2019 corresponds to the holidays.

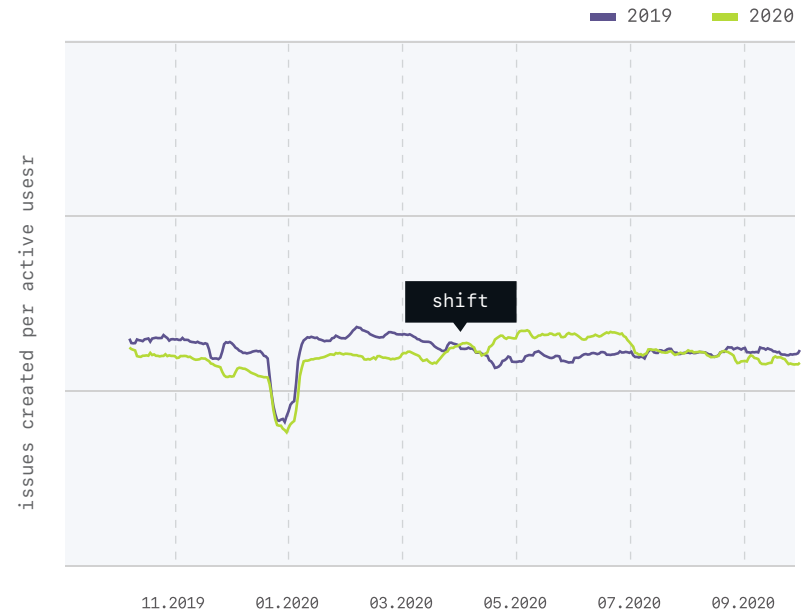

Issues created per active user, year-over-year comparison, seven-day rolling average

[2] GitHub announced free Team accounts on April 14, 2020, and analyzed the data to investigate if the increased activity in free accounts was due to a change in account type or increased activity broadly. We found that approximately half of the increase in activity is due to COVID-19, which is both a significant increase in issues for Free accounts year over year, and is noticeably different from Enterprise account activity.

Upon further investigation, we see this shift alongside an increase in issue creation rates across all repositories, with the biggest increase seen in repositories owned by Free developer and paid Team accounts. The following chart shows the volume of issues created by repository plan vs. the previous year. Note that the high and low spike in Enterprise Cloud accounts in late November is due to the US Thanksgiving holiday. We have included the chart with seven-day rolling averages to ease readability. To show the weekend dropoff in issue creation in Enterprise repositories, but no corresponding drop in others, we have included the detailed chart in the appendix.

Comparing percent increase in issue creation, we notice differences in activity among development done in the workplace (which we proxy by issues in Enterprise Cloud repositories) and other kinds of development, such as open source, hobby development, or education (which we proxy by issues in Free repositories).

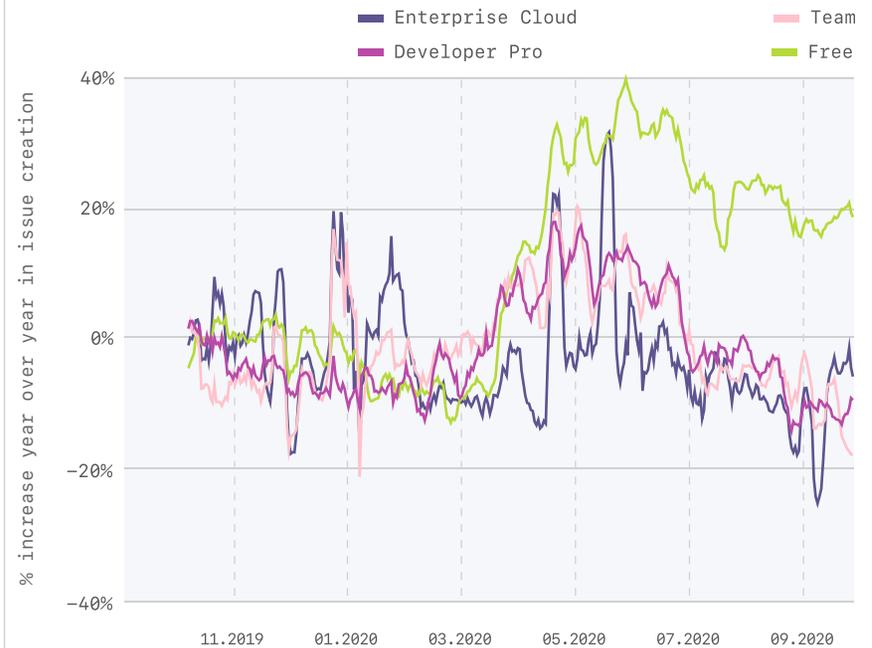

**Issues created per active user: % increase year over year by repository owner type, 7-day rolling average**

Analysis of the data combined with global events creates three distinct time periods: October through December 2019 (before COVID), January through March 2020 (early days of COVID and preparation), and April 2020 through September 2020 (when most technology workers shifted to work from home). Through this lens, we see that issue activity is different between Enterprise and Free accounts, particularly after April 2020.

**Issue creation in Enterprise Cloud and Free repositories in three time periods: Oct-Dec, Jan-Mar, Apr-Sept; changes year over year.**

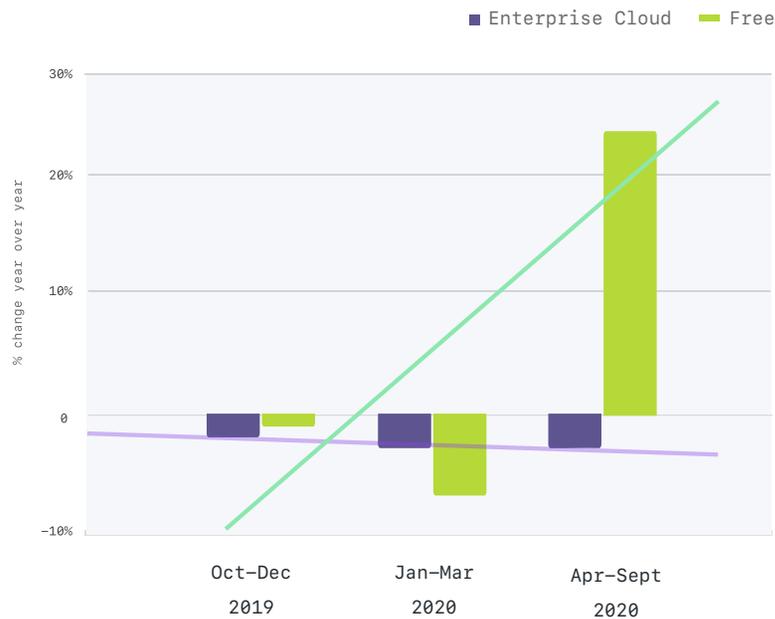

Note the trendline, and that we observe a significant change in Free repositories.

## Development stays steady

If we apply these date ranges to analyze pushes and pull requests, we see a slight increase during the year, but no significant shifts in activity.

Because pushes and pull requests are core to development activity, they don't change when work moves.

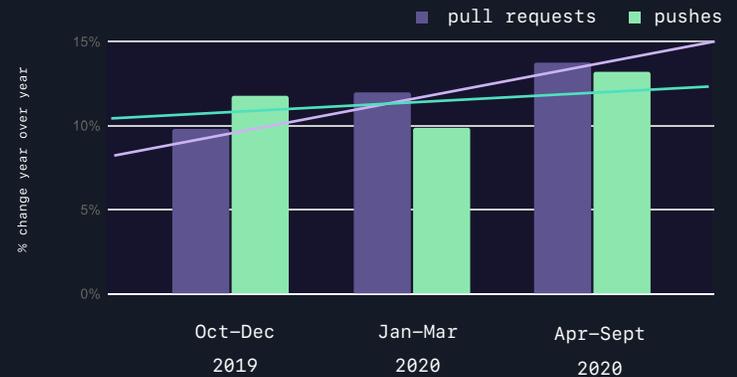

In fact, we found that even as developers were working more, the size of pushes—proxied by the number of files changed per commit—stayed roughly the same over the past year. In contrast, issues are used to track and plan work, and are more susceptible to disruptions.

**Enterprise Cloud repositories and issues**

Enterprise Cloud repositories show two primary patterns: steady year over year (with expected spikes over the holidays) until April, and then generally decreased activity with some small jumps.

We see some surges in activity for issue creation around mid-April and in May, likely as developers adjusted to working from home and coordinating development work remotely. However, we do not see issue creation rates return to their levels from the previous year. This could be because enterprise teams are more used to in-person collaboration and brainstorming sessions to kick off major initiatives, which would then be used to create issues for teams to work from.

**Free repositories and issues**

Free repositories also show two primary patterns of activity: steady or slightly reduced year over year until April. Then we see strong growth in issue creation that seems to level off in May (overall 21% average, 22% median for the period April to the present). We don't see the same strong dip in issue creation rates on the weekends in free repositories, likely because these are associated with open source, hobby, and educational work.

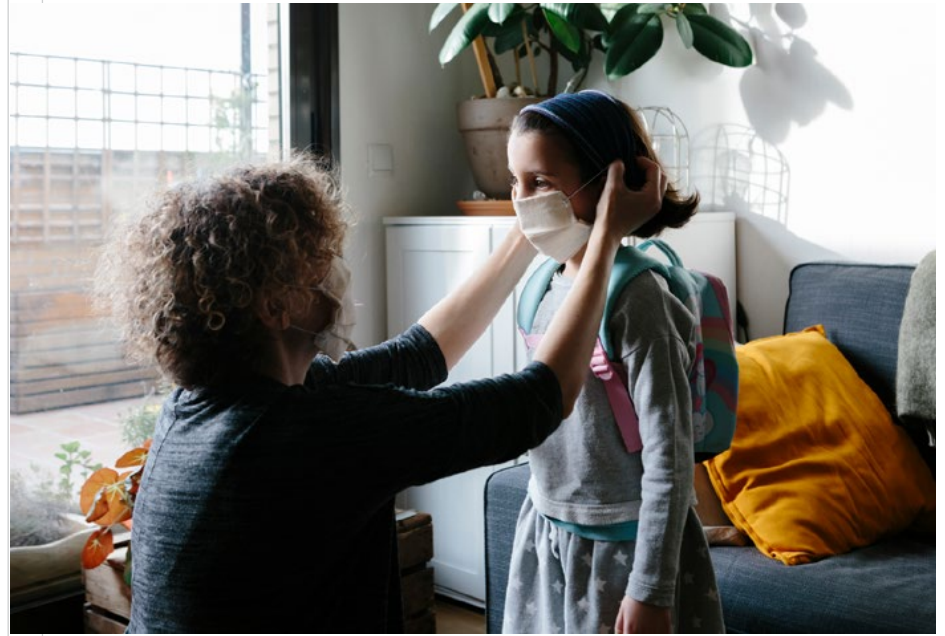

# Automation drives productivity and improves developer experience

Research shows that DevOps delivers value, but many don't know that DevOps started as a more-humane way to build software.

Teams suffered through weeks- and months-long releases. The deployments were often difficult and brittle, and teams would burn out, just to repeat it all again when the next release came out.

Several groups knew there must be a better way. By using automation and processes to improve their workflow, along with closer collaboration between teams, they were able to streamline their work. It was faster, more efficient, and best of all, created a development workflow that was sustainable: no more epic marches for deployments, and a lot less burnout.  We can use DevOps principles to make our work more sustainable:

- Leverage automation to reduce manual work and gain consistency, reliability, and efficiency.

- Use the cloud to easily provision environments so we can get to work faster and scale our services to meet demand.

- Secure our work to manage access across locations and environments.

- Build and foster a culture with strong communication and collaboration.

Using automation to deliver code quickly and reliably is important, and it's an important part of productivity. When developers can write code, get fast feedback, and then be confident their code will deploy, they can focus on tackling the next exciting problem—instead of manually deploying their code.

We looked at how open source repositories use Actions to automate their pull requests. We focused on pull requests because they are a key handoff point in the development process. By introducing automation at this stage, teams can notify others to review the pull request, and once it's reviewed, kick off tests and builds.

Across all open source repositories, once a repository starts using Actions, the time to merge a pull request decreases by 18% and the number of pull requests merged increases by 34%. Teams using automation in their workflows are accelerating their software delivery because they can merge pull requests faster, get back to coding sooner, and in turn, quickly merge more code into their project. This is a virtuous cycle, where improved software development practices create a better developer experience and continue to pay dividends.

Automation does more than accelerate software delivery. Other research shows that it also decreases errors and improves quality, which means  that developers have more time for development and innovation.

## Industry comparisons

We are often asked what the "industry standard" is for developer activity. We encourage teams to use these benchmarks to think about their own coding practices, how they compare, and how they might improve.

Developing code can be different when we are doing it at work or as a hobby, so to control for differences, we limited it to those shipping in enterprise contexts. In these cases, developers were typically committing four times per day.

**Time to merge pull request: 1.6 hours**
**Code review turnaround time: one hour**

When we break down the pull request review timeline, we see that the time to first review is typically 54 minutes, and the time from last review to merge is 12 minutes. The overall time to merge a pull request is one hour and 36 minutes, noting that adding the median values of phases within a process will not always sum. We also note that in most cases, there was only one reviewer on a pull request, so there was no time between the first review and the last, but additional reviewers could introduce delays.

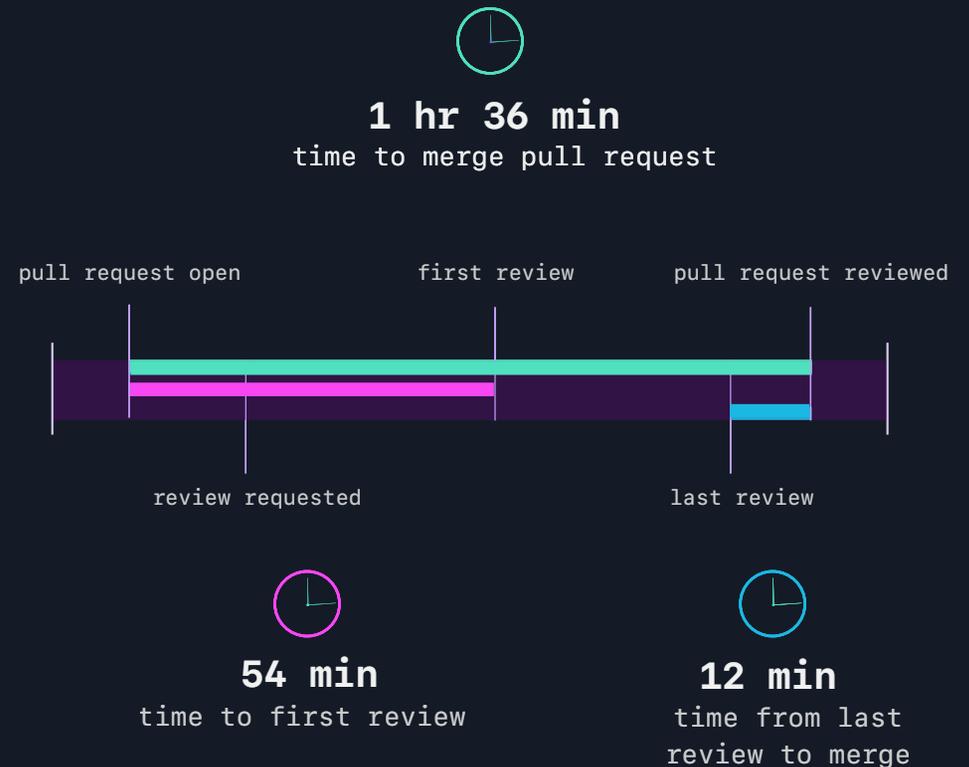

**Pull request review timeline**

1 hr 36 min
time to merge pull request

pull request open | first review | pull request reviewed
review requested | | last review

54 min
time to first review

12 min
time from last review to merge

# The new remote world

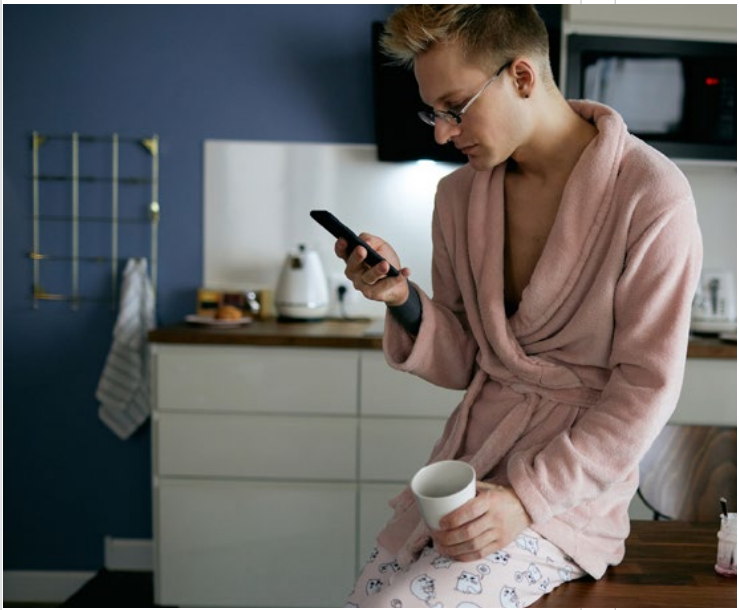

A  s we look to the future of work, this natural experiment has shown us that we are all much more successful and productive working from home than we previously thought possible. Industries such as healthcare—that just a year ago proclaimed remote work an impossibility—have found ways to deliver services in secure, reliable ways because the world and environment demanded it.

Our own analysis has shown that the number of GitHub developers has grown year over year, consistent with typical platform growth, and that individual developer activity has also increased. This is notable as it occurred even as many developers and companies shifted to a new work-from-home model and many companies pivoted to offer new features based on shifting economic and market conditions. This sustained activity through a significant shift in working conditions demonstrates that cloud-based development supports more stable and resilient development for people, teams, and organizations. Our analysis on work timing and work volume shows that developers benefit from flexibility that allows them to do their work by spreading it out.

What does this mean for our workplaces? Findings from our own research, combined with research from teams that have begun to return to work, suggests it will look like this:

- Remote and hybrid environments are likely to be the norm as teams find what works best for them.

- The longer work weeks we see now may continue, even after we return to traditional workplaces. In particular, the new "night shift" is more common.

- Providing flexible solutions so developers can create solutions that work for them makes work sustainable.

- Maintaining strong networks and collaborations are important, and these communities can remain strong through disruptions.

Work patterns over the year show us that people are working more and getting more done. This is likely the result of people using automation to accelerate their work, using good development practices, and allowing for flexibility as the lines between work and life are blurred.

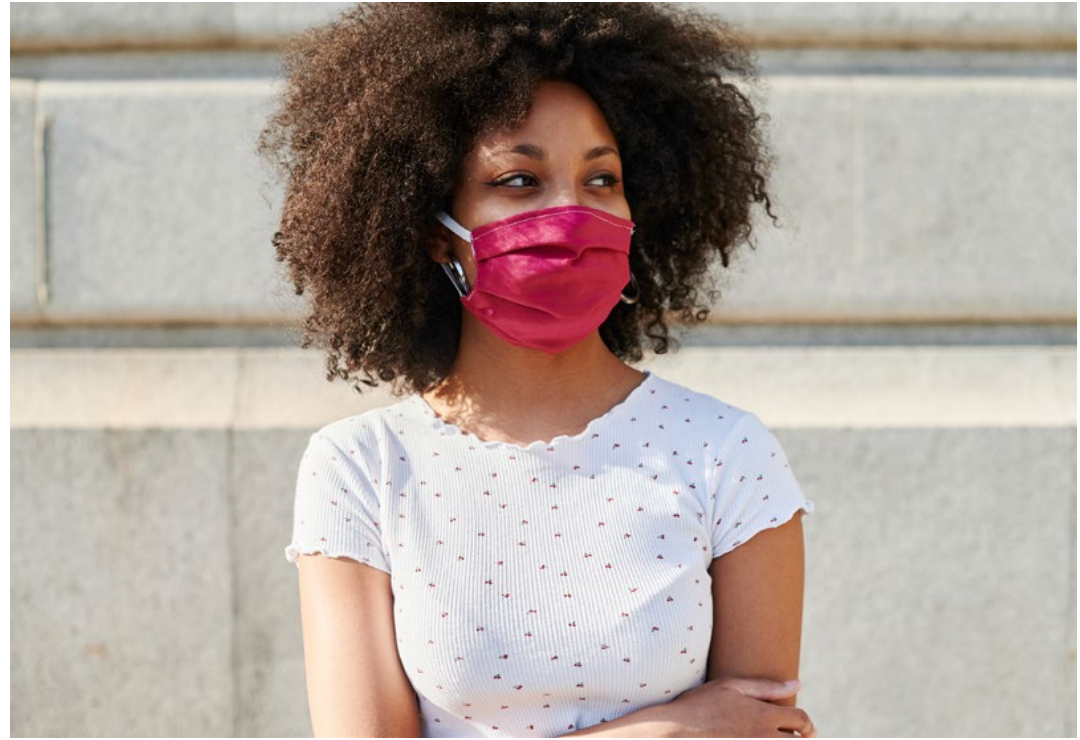

## For more insights about how we work

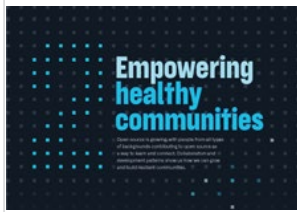

Empowering communities

**Community report** ➔

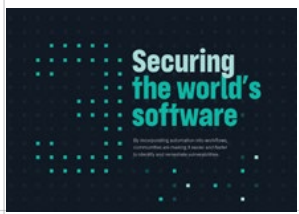

Securing software

**Security report** ➔

# We are all more resilient than we thought possible

# Glossary

**Developers**
Developers are individual accounts on GitHub, regardless of their activity.

**Location**
Country information for developers is based on their last location, where known. For organizations, we take the best-known location information either from the organization profile, or the most-common country organization members are active in. We only use location information in aggregate to look at things like trends in growth in a particular country or region. We don't look at location information granularity finer than country level.

**Open source projects**
Open source projects are public repositories with an open source license.

**Organizations**
Organization accounts represent collections of people on GitHub. These can be paid or free, big or small, businesses or nonprofits.

**Projects and repositories**
We use projects and repositories interchangeably, although we understand that sometimes a larger project can span many repositories.

**Push window**
The average minutes between a user's first and last `git push` to the primary branch of any repository, used to proxy development time.

**Work volume**
The average number of pushes done in a push window, used to proxy the amount of work done in a day.

The methodology and data used for analysis is described throughout the report

// acknowledgements

Many thanks to our data scientists, contributors, and reviewers. Each is listed alphabetically by type of contribution.

- - - - - - - - - - - - - - - - - - - - - - -

**Authors:** Nicole Forsgren

**Data scientists**: Greg Ceccarelli, Derek Jedamski, Scot Kelly, Clair Sullivan

**Reviewers:** Denae Ford, Martin Fowler

**Copyeditors:** Leah Clark,  Cheryl Coupé, Stephanie Willis

**Designers:** Siobhán Doyle, Aja Shamblee

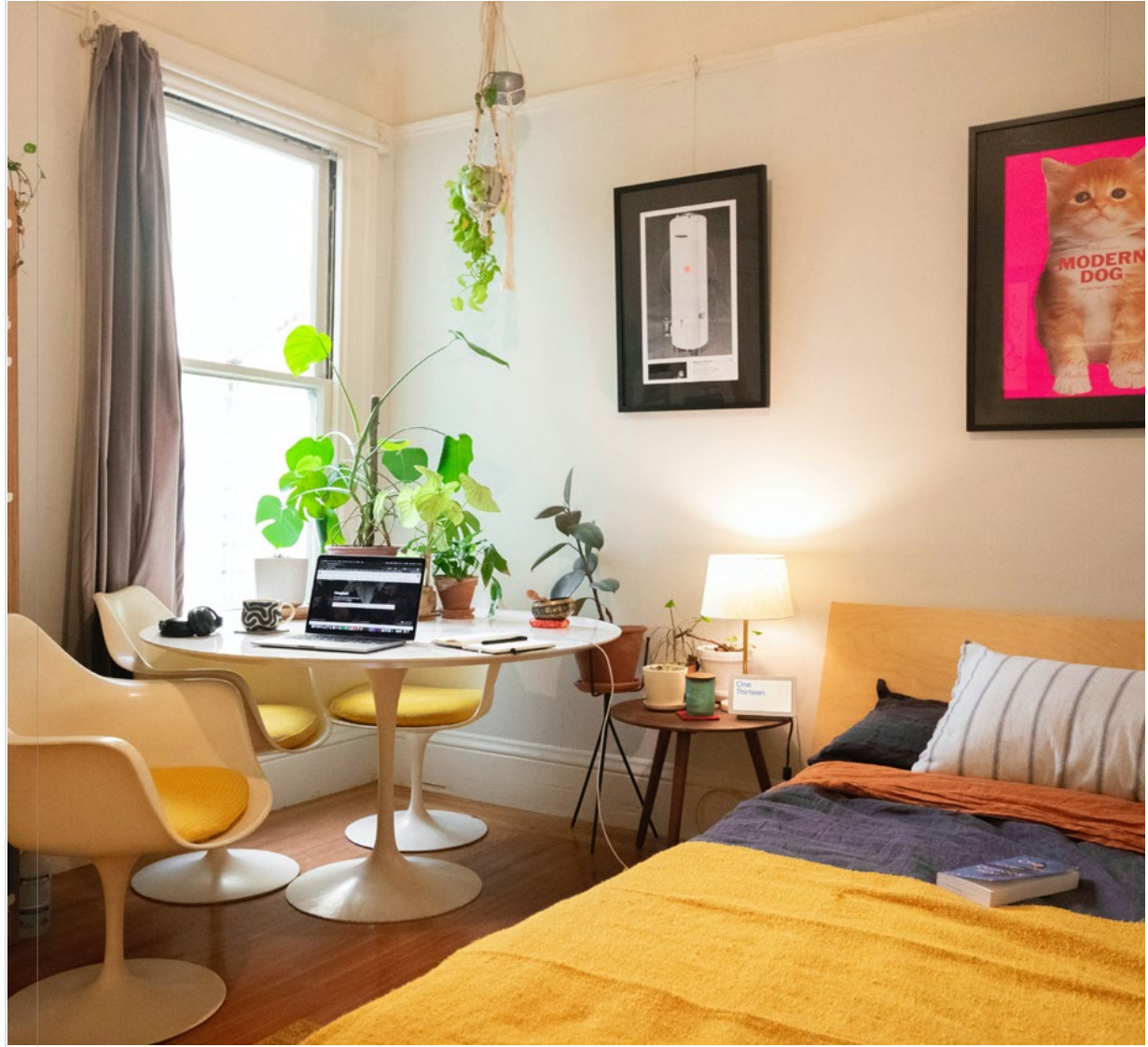

**Issues created per active user: percentage increase year over year by repository owner type**

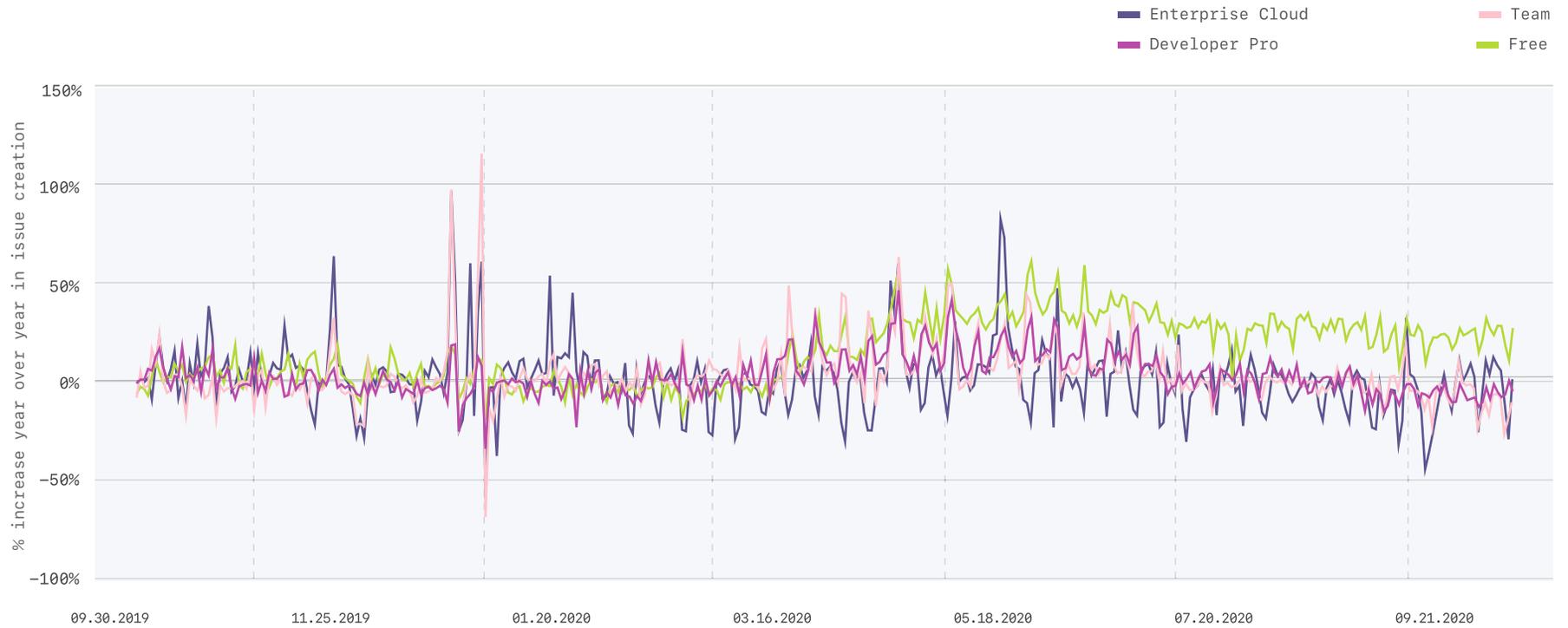

Detailed issues patterns show that creation in Enterprise repositories drops on weekends much more than others

Read more on p 28  →

**Pull requests per active user, year-over-year comparison, seven-day rolling average**

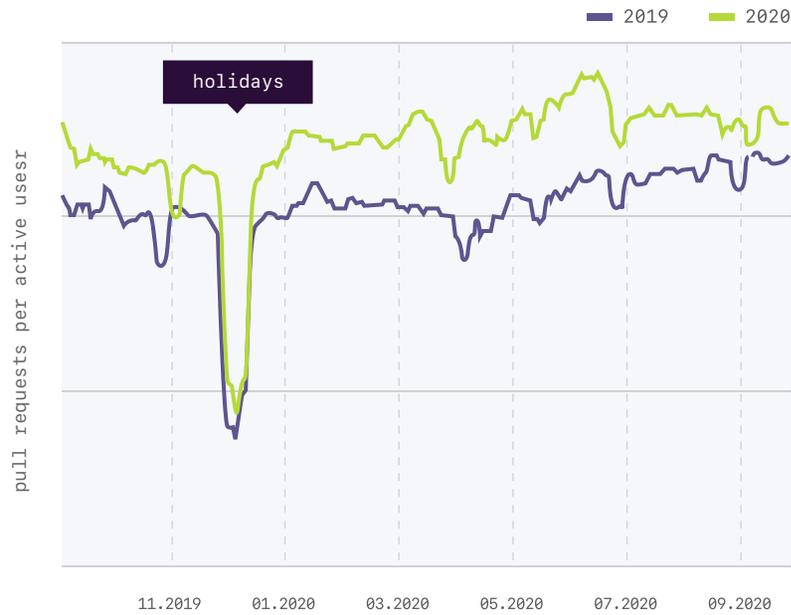

**Push volume per active user, year-over-year comparison, seven-day rolling average**

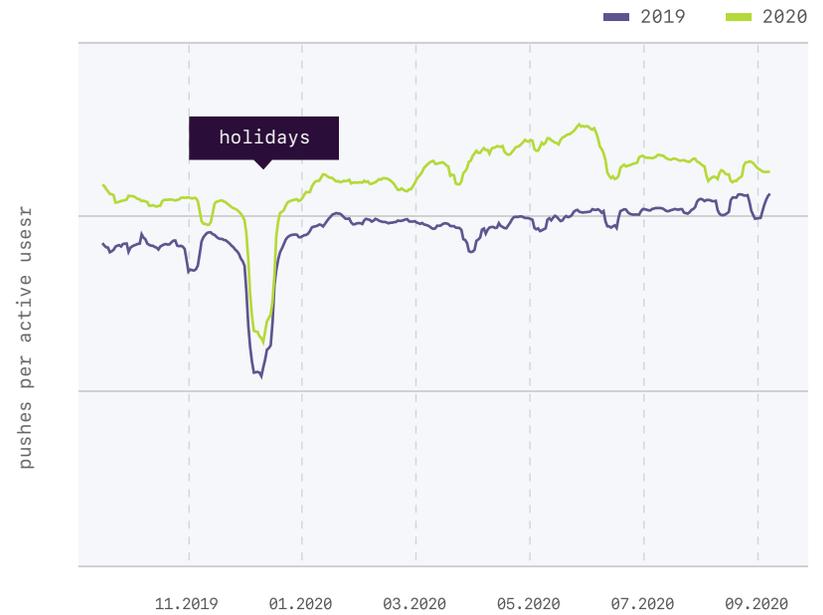

All activity charts in this report show normalized, per-person activity with a seven-day rolling average for readability